\documentclass[12pt,epsfig]{article}
\usepackage{epsfig}

\newcommand{\Pomeron}{I\!\!P}

\begin{document}

\hskip 12 cm
{RUB-TP2-11/03}

\begin{center}
{\Large {\bf
Leading twist coherent diffraction on nuclei in deep inelastic scattering 
at small $x$ and 
nuclear shadowing}}
\vskip 0.5cm
L. FRANKFURT\\
{\it Nuclear Physics Dept., School of Physics and Astronomy, Tel Aviv
  University, 69978 Tel Aviv, Israel}\\
\vskip 0.25cm 
V. GUZEY\\
{\it Institut f{\"u}r Theoretische Physik II, Ruhr-Universit{\"a}t
Bochum,
{\nobreakspace} D-44780 Bochum, Germany}\\
\vskip 0.25cm
M. STRIKMAN\\
{\it Department of Physics, the Pennsylvania State University, State
  College, PA 16802, USA}
\end{center}

{\abstract We extend the theory of leading twist nuclear shadowing to
calculate leading twist  nuclear diffractive
parton distribution functions (nDPDFs).
We observe that the quark and gluon nPDFs have different patterns of the
$A$-dependence.
 It is  found that the probability of diffraction
in the quark channel increases with $A$,
reaching
about $30\%$ at $x \sim 10^{-4}$ for $A \sim 200$, and
weakly decreases with $Q^2$. In the gluon channel, the probability
of diffraction is large for all nuclei ($\sim 40$\%
for heavy nuclei at $x \sim 10^{-4}$ and  $Q_0^2 \sim 4$ GeV$^2$), it
weakly depends on $A$  and it decreases rather fast with increasing $Q^2$ --
the probability decreases by approximately a factor of two
as $Q^2$ changes from 4 GeV$^2$ to 100 GeV$^2$.
We also find that nuclear shadowing breaks down Regge factorization of nDPDFs,
which is satisfied experimentally
in the nucleon case. All these novel effects in nDPDFs are large enough to be 
straightforwardly measured in  ultraperipheral collisions at the LHC.}

\section{Introduction}
\label{sec:intro}

It is firmly established by the HERA H1~\cite{H1:1994,H1:1997} and 
ZEUS~\cite{ZEUS:1994} experiments that inclusive diffraction constitutes 
a significant fraction, about 10\%, 
of the total cross section for deep inelastic scattering (DIS) of leptons 
on hydrogen.  The diffractive events are characterized by the absence of
hadronic activity in the detector at central rapidities. 
This QCD phenomenon 
involves both nonperturbative and perturbative aspects of the QCD dynamics,
for the discussions see~\cite{Future}, 
and it is usually referred to as scattering off  the ``Pomeron''.
Note that in the case of DIS, the dynamics  is quite different from 
that of the Pomeron pole exchange in soft hadron-hadron interactions,
so we use the concept of Pomeron in order to indicate that we
refer to the kinematics of large rapidity gaps and small momentum transfers to the recoil nucleon.

The proof of the 
 factorization theorem for hard diffraction~\cite{Collins:fact} 
enables one to describe the process in terms of $Q^2$-dependent
diffractive parton distribution functions  (DPDFs) and extract the
DPDFs from various  diffractive data.  The current
 data~\cite{H1:1994,H1:1997,ZEUS:1994}
are consistent with the dominance of leading twist in hard diffraction 
and with the dominance of the gluon DPDF over the sum of the quark  DPDFs.
The DPDFs for the scattering off unpolarized target, 
$f_{j/N}^{D(4)}(\beta,Q^2,x_{\Pomeron},t)$
depend on four variables: Bjorken $x$, virtuality $Q^2$, 
four-momentum transfer squared to the target 
$t$ and the fraction of the longitudinal momentum loss by the target $x_{\Pomeron}$.

The aim of the present paper is to 
investigate, within the leading twist approximation, nuclear dependence of
coherent (without nuclear break-up)  diffraction induced by hard probes 
and to obtain nuclear DPDFs.
This would allow to calculate the cross sections of various
diffractive DIS processes as well as of
direct photon diffraction off 
nuclear targets. Experimental studies of these processes
will be feasible in  ultraperipheral collisions  (UPC) of heavy ions at the 
LHC~\cite{LHC} and at the EIC~\cite{EIC}.

Inclusive hard diffraction in DIS off nuclei 
has been studied in a number of papers, 
see e.g.~\cite{FS96,Levin}. Unlike all previous attempts, we use 
the QCD factorization theorem for hard diffraction~\cite{Collins:fact} and
the leading twist theory of nuclear shadowing~\cite{LT1,LT2,LT3}.
This enables us, for the first time, to calculate nDPDFs, i.e. to perform the 
flavor separation. This is an essential ingredient
for the calculation of various, more 
complicated, problems such as charm production in DIS and dijet production in 
direct photon diffraction 
off nuclear targets, 
which need nDPDFs and  especially the gluon nDPDF.
The present paper complements our studies of 
the role of leading twist nuclear shadowing at small-$x$ in 
inclusive processes with nuclei.

While we assume that the QCD factorization theorem for hard diffraction 
in DIS holds and all considered effects are
leading twist effects, the studies of coherent diffraction on nuclei help 
to understand the transition to the 
regime of high parton densities. Indeed, the fraction of 
diffraction of the total cross section in DIS is a measure of how close to the 
black body regime (the regime of complete absorption of the projectile 
by the target) one is. While diffraction is approximately 10\% of the total 
cross section in DIS on hydrogen, the fraction of diffractive events 
steadily increases as one increases the atomic number $A$, 
asymptotically approaching the absolute limit of one half.

This paper is structured as follows. In Sect.~\ref{sec:main1} we 
recapitulate essential points
of the leading twist nuclear shadowing model and present the formula 
for coherent diffraction
on nuclei. The analysis and discussion of the resulting expressions 
for nuclear diffractive parton distributions are presented in  
Sect.~\ref{sec:main2}. We conclude and summarize in 
Sect.~\ref{sec:conclusions}.

\section{Leading twist nuclear shadowing and coherent diffraction on nuclei}
\label{sec:main1}

The theory of leading twist nuclear shadowing 
is based on the Gribov's relation between 
nuclear shadowing and diffraction~\cite{Gribov1969}, Collins factorization 
theorem for
hard diffraction in DIS~\cite{Collins:fact} and the QCD analysis of the 
HERA data on
hard diffraction in DIS on hydrogen~\cite{H1:1994,H1:1997}. The 
foundations of the resulting theory and predictions for nuclear parton 
distribution functions and inclusive structure 
functions can be found in Refs.~\cite{LT1,LT2,LT3}. 

The master equation for the evaluation of the shadowing correction, 
$\delta f_{j/A}$, to
the nuclear structure parton distribution functions of flavor $j$, 
$f_{j/A}=A f_{j/N}-\delta f_{j/A}$ has the form
\newpage
\begin{eqnarray}
&&\delta f_{j/A}(x,Q_0^2)=\frac{A(A-1)}{2} 16 \pi {\cal R}e \Bigg[\frac{(1-i\eta)^2}{1+\eta^2} \int d^2 b \int^{\infty}_{-\infty} dz_1 \int^{\infty}_{z_1} dz_2 \int^{x_{\Pomeron,0}}_{x} d x_{\Pomeron}  \nonumber\\
&&\times f_{j/N}^{D(4)}(\beta,Q_0^2,x_{\Pomeron},t_{{\rm min}}) \rho_A(b,z_1) \rho_A(b,z_2) e^{i x_{\Pomeron} m_N (z_1-z_2)} e^{-(A/2)(1-i\eta)\sigma_{{\rm eff}}^j \int_{z_1}^{z_2} dz \rho_A(b,z)} \Bigg] \,,
\label{eq:master} 
\end{eqnarray} 
with $\eta$ the ratio of the real to imaginary parts of the diffractive 
scattering amplitude; $z_1$, $z_2$ and $\vec{b}$ the longitudinal (in 
the direction of the incoming virtual photon) and transverse coordinates 
of the nucleons involved (defined with respect to the nuclear center);     
$\beta$, $x_{\Pomeron}$ and $t$ the usual kinematic variables used in 
diffraction; $\beta=x/x_{\Pomeron}$; $t_{{\rm min}} \approx 0$;
$\rho_A(b,z_i)$ the nucleon distribution in the target nucleus.
The upper limit of integration, $x_{\Pomeron,0}$ is a cut-off parameter, 
which equals 0.1 for quarks and 0.03 for gluons.
The effective cross section, $\sigma_{{\rm eff}}^j$,
 is expressed through the nucleon DPDFs as 
(see Ref.~\cite{LT3} for the detailed discussion and numerical estimates)
\begin{equation}
\sigma_{{\rm eff}}^j(x,Q_0^2)=\frac{16 \pi}{f_{j/N}(x,Q_0^2)(1+\eta^2)}\int_x^{x_{\Pomeron,0}}d x_{\Pomeron}  f_{j/N}^{D(4)}(\beta,Q_0^2,x_{\Pomeron},t)\big|_{t=t_{{\rm min}}} \,.
\label{eq:sigma} 
\end{equation}
Equation~(\ref{eq:master}) serves to define the input nuclear PDFs at the 
initial scale $Q_0^2$,  $Q_0^2=4$ GeV$^2$ in our analysis. Nuclear PDFs at larger
scales $Q^2$ are obtained using the NLO QCD evolution equations.

In  Eq.~(\ref{eq:master}), the interaction with two nucleons is calculated 
in a model-independent way.  The only source of model-dependence is
due to the approximation of the 
interaction with three and more nucleons by the attenuation factor
$e^{-(A/2)(1-i\eta)\sigma_{{\rm eff}}^j \int_{z_1}^{z_2} dz \rho_A(b,z)}$,
 which involves $\sigma_{{\rm eff}}^j$, the
 rescattering cross section given by Eq.~(\ref{eq:sigma}). 
While this quasi-eikonal approximation is expected to be valid at
$Q_0^2=4$ GeV$^2$, it becomes progressively worse with increasing $Q^2$.
The reason for this is that the eikonal approximation conserves the number
of bare particles and thus contradicts QCD evolution. As a result, one
obtains a wrong, higher twist, $Q^2$-dependence of nuclear shadowing
in the processes dominated by small partonic configurations of the 
incoming virtual photon. Only at low $Q^2$ scales, where the effects of 
QCD evolution are not very important, can one justify the use of the 
eikonal and quasi-eikonal approximations.
This means that 
Eq.~(\ref{eq:master}) should be used only at the initial scale 
$Q_0^2=4$ GeV$^2$.

The generalization to the case of coherent diffraction in DIS on nuclei 
is rather straightforward, and
it follows closely the case of the vector meson diffraction,
see e.g.~\cite{BPY}. The nuclear diffractive parton distribution of flavor $j$
can be  presented in the form
\begin{eqnarray}
f_{j/A}^{D(3)}(x,Q_0^2,x_{\Pomeron})&=&\frac{A^2}{4} 16 \pi f_{j/N}^{D(4)}(x,Q_0^2,x_{\Pomeron},t_{{\rm min}}) \int d^2 b \nonumber\\
&&\times \left| \int^{\infty}_{-\infty} dz e^{i x_{\Pomeron} m_N z} e^{\sigma_{{\rm eff}}^jA (1-i \eta)/2 \int_{z}^{\infty} \rho_A(b,z^{\prime})} \rho_A(b,z)\right|^2 \,.
\label{eq:masterD}
\end{eqnarray}
The superscripts $(3)$ and $(4)$ denote the dependence on three and four 
variables, respectively. We present
our Eq.~(\ref{eq:masterD}) for the $t$-integrated nuclear DPDFs since 
it is more compact and since it is not feasible to 
measure $t$ in diffraction off nuclei in the collider experiments.
In deriving Eq.~(\ref{eq:masterD}) we neglected a possible $\beta$-dependence
of $\sigma_{\rm {eff}}^j(x,Q^2)$ 
in the exponential factor and substituted $\sigma_{\rm {eff}}^j$ by its 
average value. Since the total probability of
diffraction changes rather weakly with $\sigma_{\rm {eff}}^j$, see 
e.g.~\cite{Krakow}, this seems a reasonable first approximation.
At  the same time, in the region of small $\beta$ and small $x$ 
corresponding to the 
triple Pomeron kinematics for  soft inelastic diffraction, 
we expect a significant suppression of  diffraction
as compared to the quasi-eikonal approximation of 
Eq.~(\ref{eq:masterD}) for $Q^2 \sim Q_0^2$,
 see the discussion in the end of the section.

One should note that the large momentum transfer $Q^2$, which is necessary for the 
applicability of the QCD factorization theorem, does not preclude the existence
of coherent nuclear diffraction. Indeed, at high energies, the minimal momentum tranfer to
the nucleus $t_{{\rm min}}$ is small, $t_{{\rm min}} \approx x_{Bj}^2 M_A^2$, which
makes it possible for nucleus to stay intact (or diffract into low mass excited states). 
In practice, coherent nuclear diffraction can be identified by its distinctly sharp
$t$-dependence in the forward direction (forward diffractive peak), which originates
from the factor $\left(F_A(t)\right)^2$ where $F_A(t)$ is the nuclear form factor.

Assuming the exponential $t$-dependence of inclusive diffraction on 
free nucleons and using that
$t_{{\rm min}} \approx 0$,
we obtain
\begin{equation}
f_{j/N}^{D(4)}(x,Q_0^2,x_{\Pomeron},t=0)=B_j \, f_{j/N}^{D(3)}(x,Q_0^2,x_{\Pomeron}) \,,
\label{eq:slope}
\end{equation}
where $B_j$ is a slope. In our analysis, we use $B_j=7.2$ GeV$^{-2}$ for quarks and
$B_j=6+0.25 \ln(10^{-3}/x)$ GeV$^{-2}$ for gluons, which are rather close numerically.

 For sufficiently small values of  
$x_{\Pomeron}$, $x_{\Pomeron} \leq 0.01$, 
the H1 data~\cite{H1:1994,H1:1997} and ZEUS data~\cite{ZEUS:1994}
can be fitted reasonably well using a factorized approximation
first suggested within the picture of  soft mechanism of diffractive process
by Ingelman and Schlein~\cite{IS}. 
In this approach, 
DPDFs can be presented 
as a product of a factor depending only on  $t$ and $x_{\Pomeron}$
(Pomeron flux) and a factor depending only on $\beta=x/x_{\Pomeron}$ and  
$Q^2$ (which is often referred to as the DPDF of the Pomeron) 
\begin{equation}
f_{j/N}^{D(3)}(x,Q_0^2,x_{\Pomeron})=f_{\Pomeron /p}(x_{\Pomeron}) \, f_{j/ \Pomeron}(\beta=x/ x_{\Pomeron},Q_0^2) \,,
\label{eq:Regge}
\end{equation}
where $f_{\Pomeron /p}$ is the so-called Pomeron flux and $f_{j/\Pomeron}$
 is the parton 
distribution function of the Pomeron.

Note that the QCD fits to the diffractive data lead to  
$\alpha_{\Pomeron}(0)$ for the effective Pomeron trajectory,
which is somewhat larger 
than the one for the effective soft Pomeron trajectory. 
This is likely due to a  different
interplay of soft and semihard physics in 
hard diffraction at the $Q_0^2$ scale and a different
role of  screening compared to soft interactions.
Hence, it is likely that a violation of the factorization approximation 
will be be observed once the data are more accurate.

 The final expression for $f_{j/A}^{D(3)}$ takes the form
\begin{eqnarray}
f_{j/A}^{D(3)}(x,Q_0^2,x_{\Pomeron})&=&\frac{A^2}{4} 16 \pi B_j f_{\Pomeron /p}(x_{\Pomeron}) \, f_{j/ \Pomeron}(\beta=x/ x_{\Pomeron},Q_0^2)
 \int d^2 b \nonumber\\
&&\times \left| \int^{\infty}_{-\infty} dz e^{i x_{\Pomeron} m_N z} e^{\sigma_{{\rm eff}}^j A (1-i \eta)/2 \int_{z}^{\infty} \rho_A(b,z^{\prime})} \rho_A(b,z)\right|^2 \,.
\label{eq:masterD2}
\end{eqnarray}
One immediately sees from Eq.~(\ref{eq:masterD2}) that  
the factorization approximation is not valid for 
nuclear diffractive parton distributions,
even if it is valid for the nucleon
case:
 At fixed $x_{\Pomeron}$, the right hand side of
Eq.~(\ref{eq:masterD2}) depends not only on $\beta$ but also on the 
Bjorken $x$ since the screening factor 
is given by the exponential factor containing
$\sigma_{{\rm eff}}^j$, which 
is a function of $x$. In addition, the right hand side of
Eq.~(\ref{eq:masterD2}) depends on the atomic mass 
number $A$ since the effect of  nuclear
shadowing increases with increasing $A$.

The aforementioned breakdown of the factorization approximation 
is a result of the  increase of the nuclear shadowing effects both
 with the increase of incident energy and with the increase of the 
atomic number. This precludes the possibility of a scenario 
offered in Ref.~\cite{Krasny},
where coherent diffraction in DIS on nucleon and nuclear targets
is provided by the same universal diffractive PDFs --``a universal Pomeron''.

It is also worth noting that the approximation which we use at 
$Q_0^2$ in order 
to take into account multiple rescatterings, corresponds essentially to
treating diffraction as  superposition of  elastic scattering of different 
components of the photon
wave function off the nucleus. This is a reasonable approximation 
for the configurations
 with masses comparable to
$Q^2$. As one approaches the $\beta \ll 1$ limit (which corresponds to $M_X^2 \gg Q^2$,
 one approaches the limit analogous
to the soft triple Pomeron limit, in which case  diffraction
 off nuclei is strongly
 suppressed as compared to the elastic scattering,
 see e.g.~\cite{FS96,Kaidalov:Krakow}.
This effect should be even stronger in our case of DIS since 
$\sigma_{\rm eff}^j$ increases with the decrease of $\beta$.
Hence, we somewhat overestimate diffraction at small $\beta$ and at  
relatively small $Q_0^2$ scale, see Figs.~\ref{fig:ca40_beta_new} and 
\ref{fig:pb_beta_new}. At larger $Q^2$, diffraction
at small $\beta$ is dominated by the QCD evolution from 
$\beta \geq 0.1$ at $Q_0^2$ and, hence,
the  accuracy of our approximation improves.
Hence, in the numerical studies, we neglect the effect of the small-$\beta$
suppression.

\section{Numerical results}
\label{sec:main2}

In our analysis of Eq.~(\ref{eq:masterD2}), we used the 1994 
H1 fit~\cite{H1:1994} for $f_{j/ \Pomeron}$, where the
 gluon distribution is decreased by the factor 0.75. This change 
seems to be required by 
the more recent analysis of the 1997 H1 data on inclusive diffraction 
on hydrogen~\cite{H1:1997}.
 
\begin{figure}[h]
\begin{center}
\epsfig{file=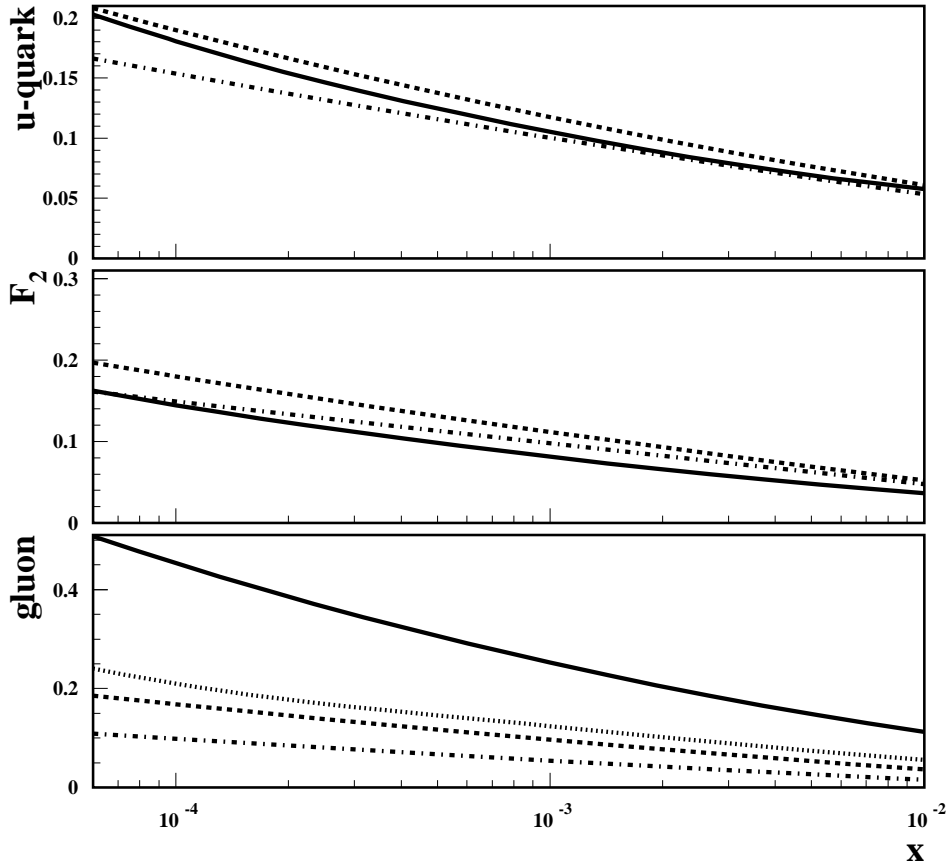,width=13cm,height=13cm}
\vskip 0cm
\caption{The ratios $f_{j/N}^{D(2)}/f_{j/N}$ for the $u$-quarks and gluons and 
NLO $F_{2N}^{D(2)}/F_{2N}$. The solid curves correspond to $Q=2$ GeV; the dashed 
curves correspond to $Q=10$ GeV; the dot-dashed curves correspond to $Q=100$ GeV.
In addition, for the gluons the dotted curve correspond to $Q=5$ GeV.
}
\label{fig:nucleon}
\end{center}
\end{figure}

In order to have an idea about the magnitude of diffraction in DIS on 
hydrogen, the ratios
 $R_{j/N} \equiv f_{j/N}^{D(2)}/f_{j/N}$ for  $u$-quarks and gluons and 
$F_{2N}^{D(2)}/F_{2N}$ are presented in Fig.~\ref{fig:nucleon}
as functions of Bjorken $x$. Note that by definition
\begin{equation}
f_{j/N}^{D(2)}(x,Q^2)=\int_{x}^{x_{\Pomeron,0}} d x_{\Pomeron} \,  f_{j/N}^{D(3)}(x,Q^2,x_{\Pomeron}) \,.
\label{eq:d2}
\end{equation}
These ratios give the probability of 
diffraction in the processes  dominated by the coupling 
of a hard probe to a quark or gluon, respectively.
For the hard process with a 
specific trigger, the probability of  diffraction maybe close to 
$R_{q/N}$, such as the measurement of the diffractive structure function 
$F_{2N}^{D}$, or to
 $R_{g/N}$, such as the $b$-quark production. 
Alternatively, the probability of
diffraction can have an intermediate value between $R_{q/N}$ and $R_{g/N}$,
such as in the $s$-quark production.
In Fig.~\ref{fig:nucleon}, the solid curves correspond to $Q=2$ GeV; the 
dashed curves correspond to $Q=10$ GeV; the dot-dashed curves correspond to $Q=100$ GeV.
Since the $Q^2$ dependence of $g_{N}^{D(2)}/g_{N}$ is rather strong, we also show the 
ratio $g_{N}^{D(2)}/g_{N}$ at $Q=5$ GeV (dotted curve). 
One sees from Fig.~\ref{fig:nucleon} that in the quark channel, 
diffraction constitutes 15-20\%
of the total cross section, while in the gluon channel diffraction 
is significantly larger~\cite{LT1}. This
 is  likely to be related to a larger cross 
section of the interaction of the gluon color dipole 
(in the {\bf 8x8} representation) as compared to the triplet
quark-antiquark dipole.

The absolute upper limit of the gluon distribution, 
$g_{N}^{D(2)}/g_{N}=1/2$ is reached at $x=6 \times 10^{-5}$ and 
$Q=2$ GeV. Since we prefer to stay away from modeling the kinematics, 
where taming
of the increase of the diffractive parton distributions becomes necessary,
we will consider the limited range of Bjorken $x$, $x > 6 \times 10^{-5}$, 
in this paper.

\begin{figure}[h]
\begin{center}
\epsfig{file=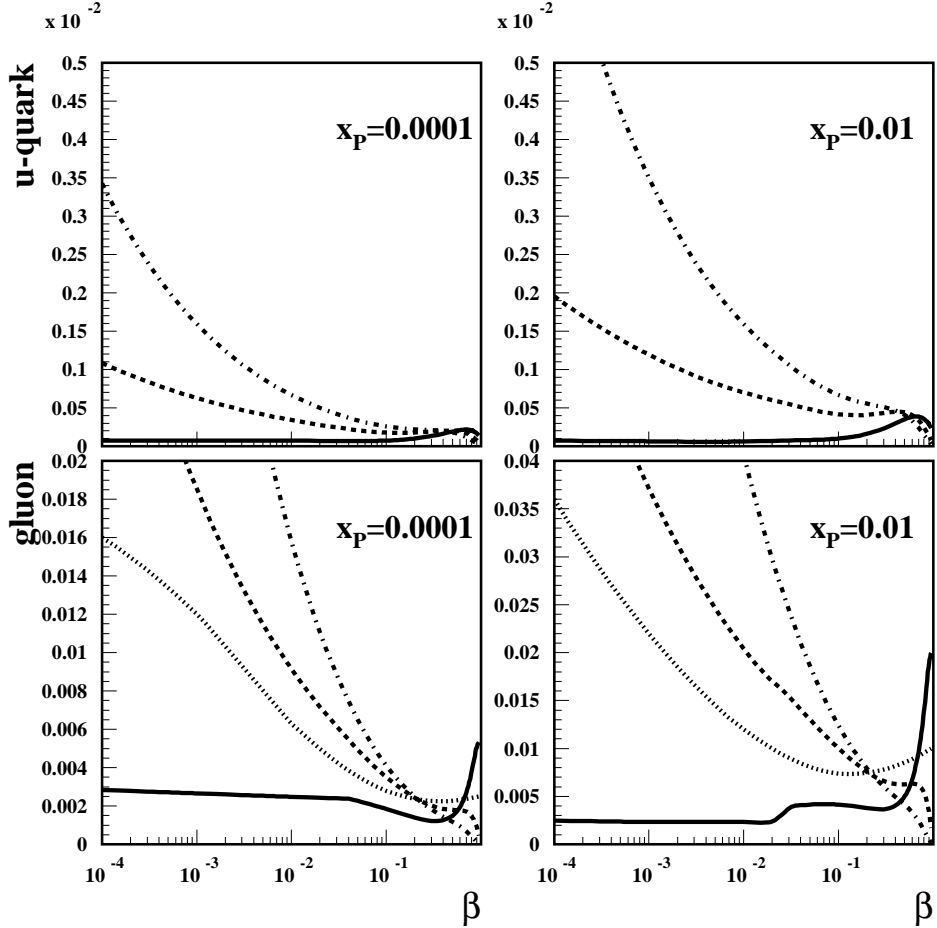,width=13cm,height=13cm}
\vskip 0cm
\caption{The $u$-quark and gluon nuclear ($^{40}$Ca) diffractive parton distribution as a 
function of $\beta$ at two fixed values of $x_{\Pomeron}$. 
The solid curves correspond to $Q=2$ GeV; the dashed curves correspond to $Q=10$ GeV; 
the dot-dashed curves correspond to $Q=100$ GeV.
In addition, for the gluons the dotted curve correspond to $Q=5$ GeV.
}
\label{fig:ca40_beta_new}
\end{center}
\end{figure}

\begin{figure}[h]
\begin{center}
\epsfig{file=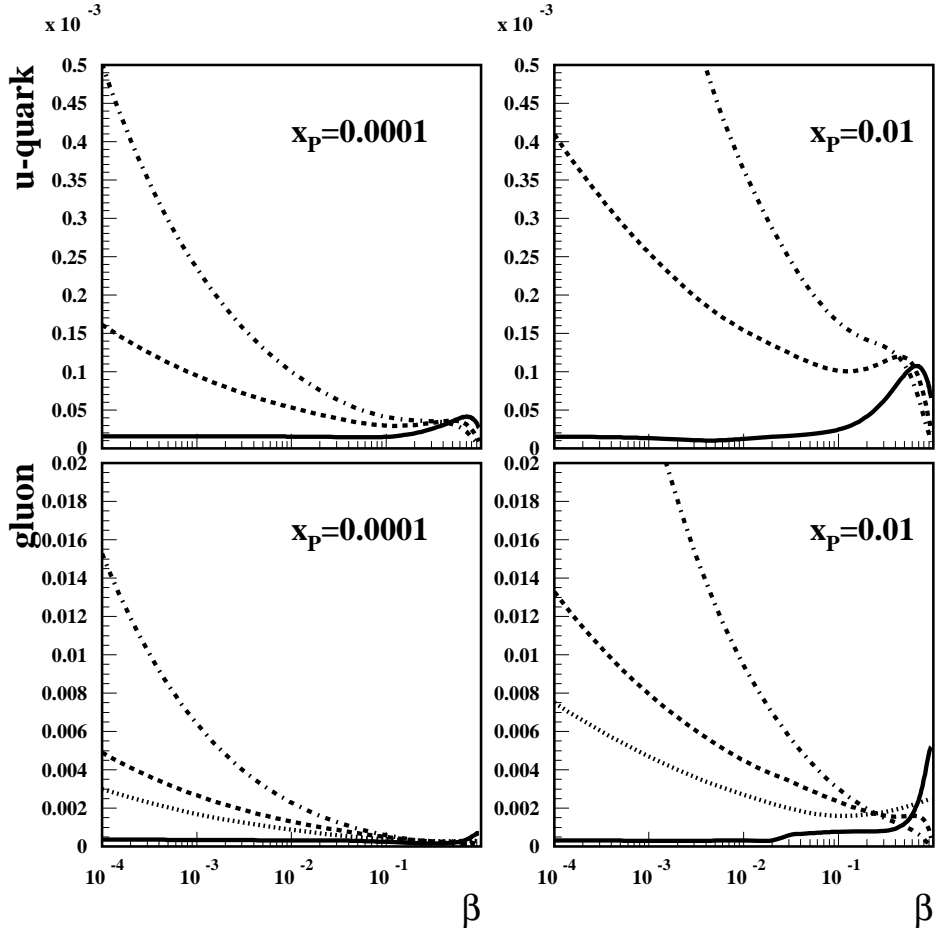,width=13cm,height=13cm}
\vskip 0cm
\caption{The $u$-quark and gluon nuclear ($^{208}$Pb) diffractive parton distribution 
as a function of $\beta$ at 
two fixed values of $x_{\Pomeron}$. 
The solid curves correspond to $Q=2$ GeV; the dashed curves correspond to $Q=10$ GeV; 
the dot-dashed curves correspond to $Q=100$ GeV.
In addition, for the gluons the dotted curve correspond to $Q=5$ GeV.
}
\label{fig:pb_beta_new}
\end{center}
\end{figure}

Next it is natural to analyze how the ratios presented in Fig.~\ref{fig:nucleon}
change when hydrogen is replaced by a  nuclear target. This can be done in two steps.
First, acting in the spirit of the QCD factorization theorem for hard diffraction in DIS,
 Eq.~(\ref{eq:masterD2}) is used to define the input for DGLAP evolution at fixed $x_{\Pomeron}$.
Subsequent QCD evolution enables us to determine $f_{j/A}^{D(3)}(x,Q^2,x_{\Pomeron})$
as a function of $\beta=x/x_{\Pomeron}$ and $Q^2$ at all fixed  $x_{\Pomeron}$.
An example of these results in presented in Figs.~\ref{fig:ca40_beta_new} and \ref{fig:pb_beta_new}
 for the nuclear targets of $^{40}$Ca and $^{208}$Pb. 
The $u$-quark and gluon nuclear diffractive parton distributions $f_{j/A}^{D(3)}(x,Q^2,x_{\Pomeron})$
(arbitrary absolute normalization) are presented as functions of $\beta=x/x_{\Pomeron}$ 
at two fixed values of $x_{\Pomeron}=10^{-4}$ and $x_{\Pomeron}=10^{-2}$.
The solid curves correspond to $Q=2$ GeV; the dashed curves correspond to $Q=10$ GeV; 
the dot-dashed curves correspond to $Q=100$ GeV;
the dotted curves correspond to $Q=5$ GeV.
 Different shapes and sizes of 
$f_{j/A}^{D(3)}(x,Q^2,x_{\Pomeron})$ at $x_{\Pomeron}=10^{-4}$ and $x_{\Pomeron}=10^{-2}$
 clearly demonstrate violation of 
the factorization approximation 
for nuclear DPDFs.

\begin{figure}[h]
\begin{center}
\epsfig{file=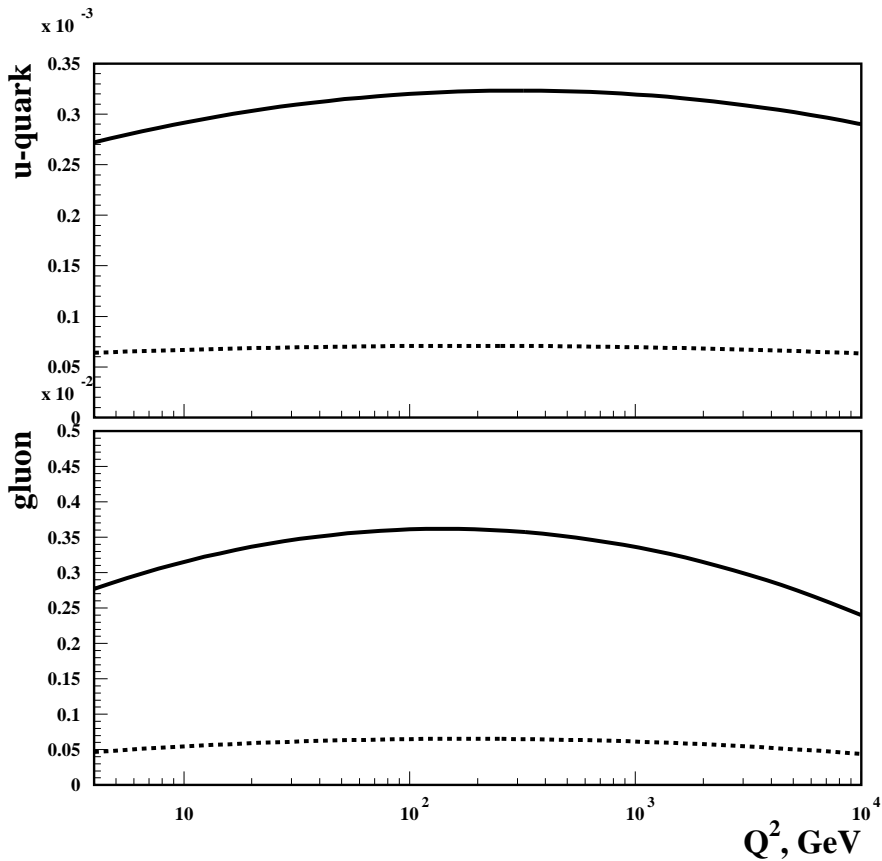,width=12cm,height=12cm}
\vskip 0cm
\caption{The $u$-quark and gluon nuclear PDFs as functions of $Q^2$ at $\beta=0.5$ and 
small $x_{\Pomeron}=10^{-3}$. The solid curves correspond to $^{40}$Ca; the dotted 
curves correspond to $^{208}$Pb.}
\label{fig:largebeta}
\end{center}
\end{figure}

Another characteristic feature of nuclear 
DPDFs is that, like in the case of the
 free proton target,
the gluon distribution is significantly larger than the quark distribution.
However, we point out that the ratio of the quark to the gluon DPDF is
 significantly larger
in the nuclear case because of a faster increase of the quark nDPDF with the
atomic number $A$.
Also, similarly to the free proton case, scaling violations
 of nDPDFs at large $\beta$ 
are rather
insignificant. This point is exemplified in Fig.~\ref{fig:largebeta} where we 
plot
the $u$-quark and gluon nDPDFs as functions of $Q^2$ at fixed large $\beta=0.5$ and 
small $x_{\Pomeron}=10^{-3}$. One readily sees from Fig.~\ref{fig:largebeta} that 
QCD evolution in $\ln Q^2$ is weak.

\begin{figure}[h]
\begin{center}
\epsfig{file=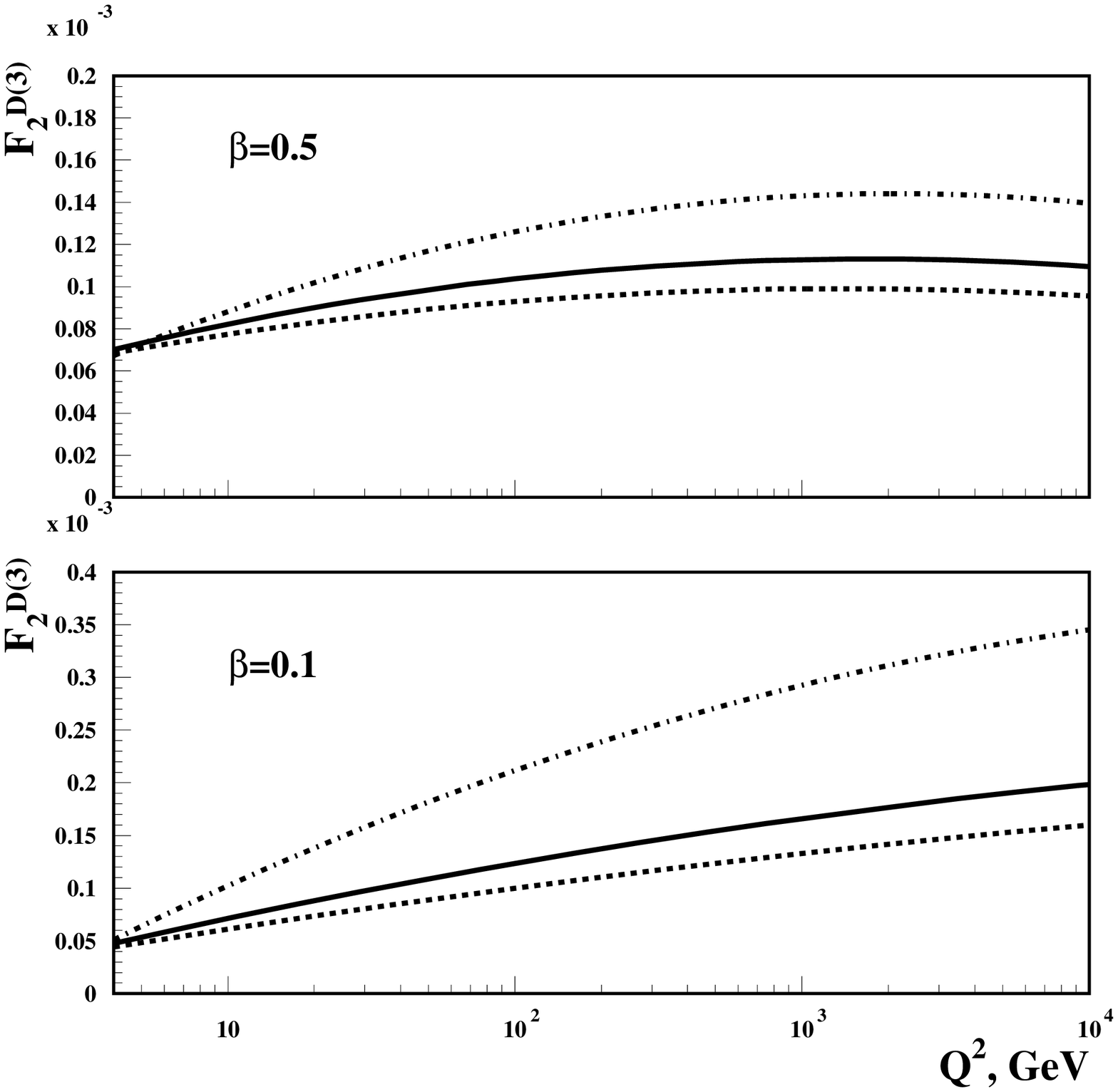,width=12cm,height=12cm}
\vskip 0cm
\caption{The diffractive structure function $F_2^{D(3)}$ 
as a function of $Q^2$ at $\beta=0.5$ and $\beta=0.1$ and 
small $x_{\Pomeron}=10^{-3}$. The solid curves correspond to $^{40}$Ca; the dotted 
curves correspond to $^{208}$Pb; the dot-dashed curves correspond to the free 
nucleon.}
\label{fig:largebeta2}
\end{center}
\end{figure}

The difference in large-$\beta$ scaling violations of the structure 
function $F_2^{D(3)}$ in the nuclear and nucleon case is presented in 
Fig.~\ref{fig:largebeta2}. At $x_{\Pomeron}=10^{-3}$ and two values of $\beta$,
$\beta=0.5$ and $\beta=0.1$, $F_2^{D(3)}$ is plotted as a function of $Q^2$ for
$^{40}$Ca (solid curves), $^{208}$Pb (dashed curves) and free nucleon 
(dot-dashed curves). The curves are normalized to coincide at the lowest $Q^2=4$ GeV$^2$.
One can readily observe from Fig.~\ref{fig:largebeta2}
that scaling violations are largest for the free nucleon and that scaling violations decrease 
as one increases $A$.

\begin{figure}[h]
\begin{center}
\epsfig{file=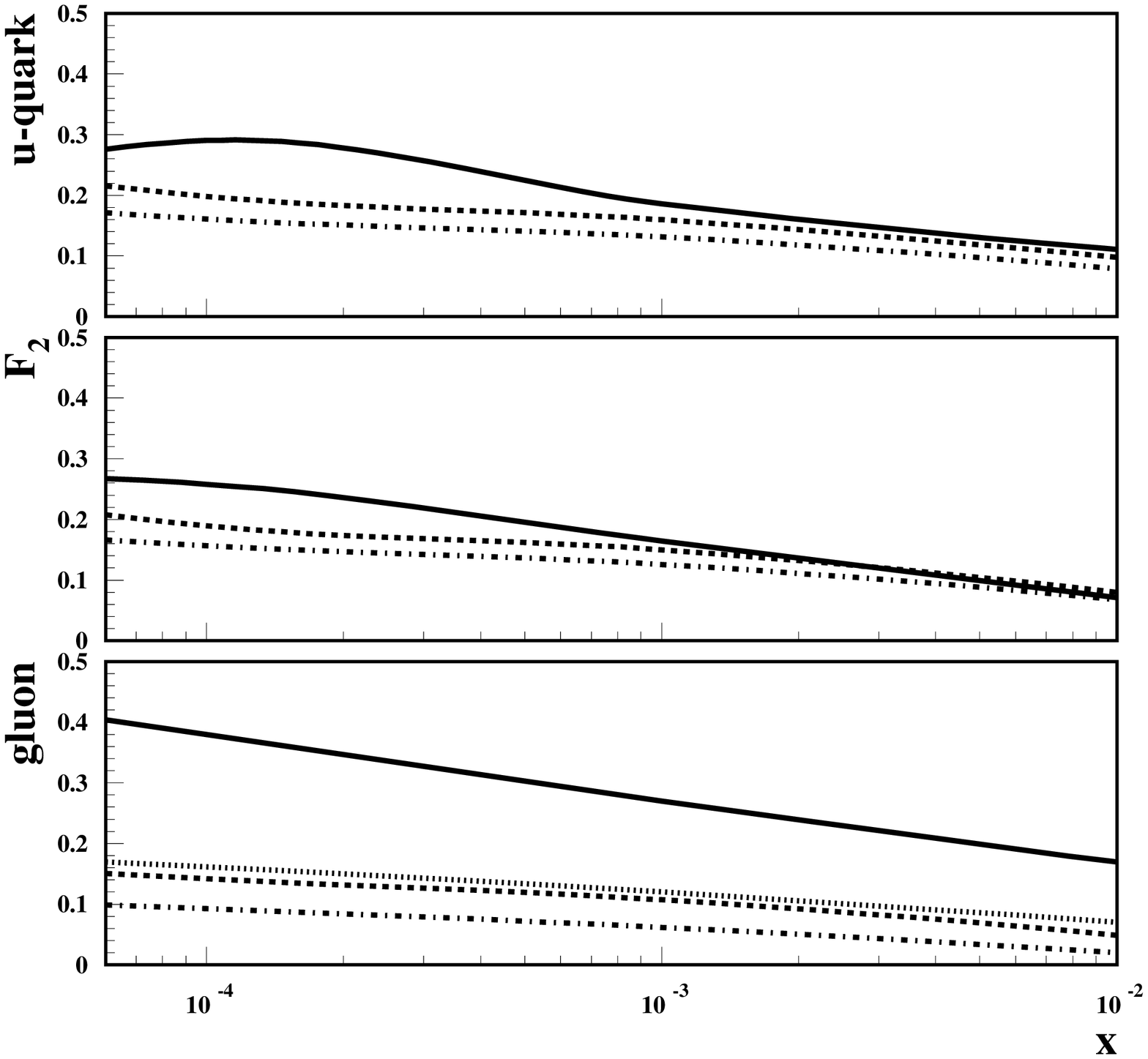,width=13cm,height=13cm}
\vskip 0cm
\caption{The ratios $f_{j/A}^{D(2)}/f_{j/A}$ for the $u$-quarks and gluons and 
NLO $F_{2A}^{D(2)}/F_{2A}$ for $^{40}$Ca. The solid curves correspond to
 $Q=2$ GeV; the dashed 
curves correspond to $Q=10$ GeV; the dot-dashed curves correspond to $Q=100$ GeV.
In addition, for the gluons the dotted curve correspond to $Q=5$ GeV.
}
\label{fig:ca40}
\end{center}
\end{figure}

\begin{figure}[h]
\begin{center}
\epsfig{file=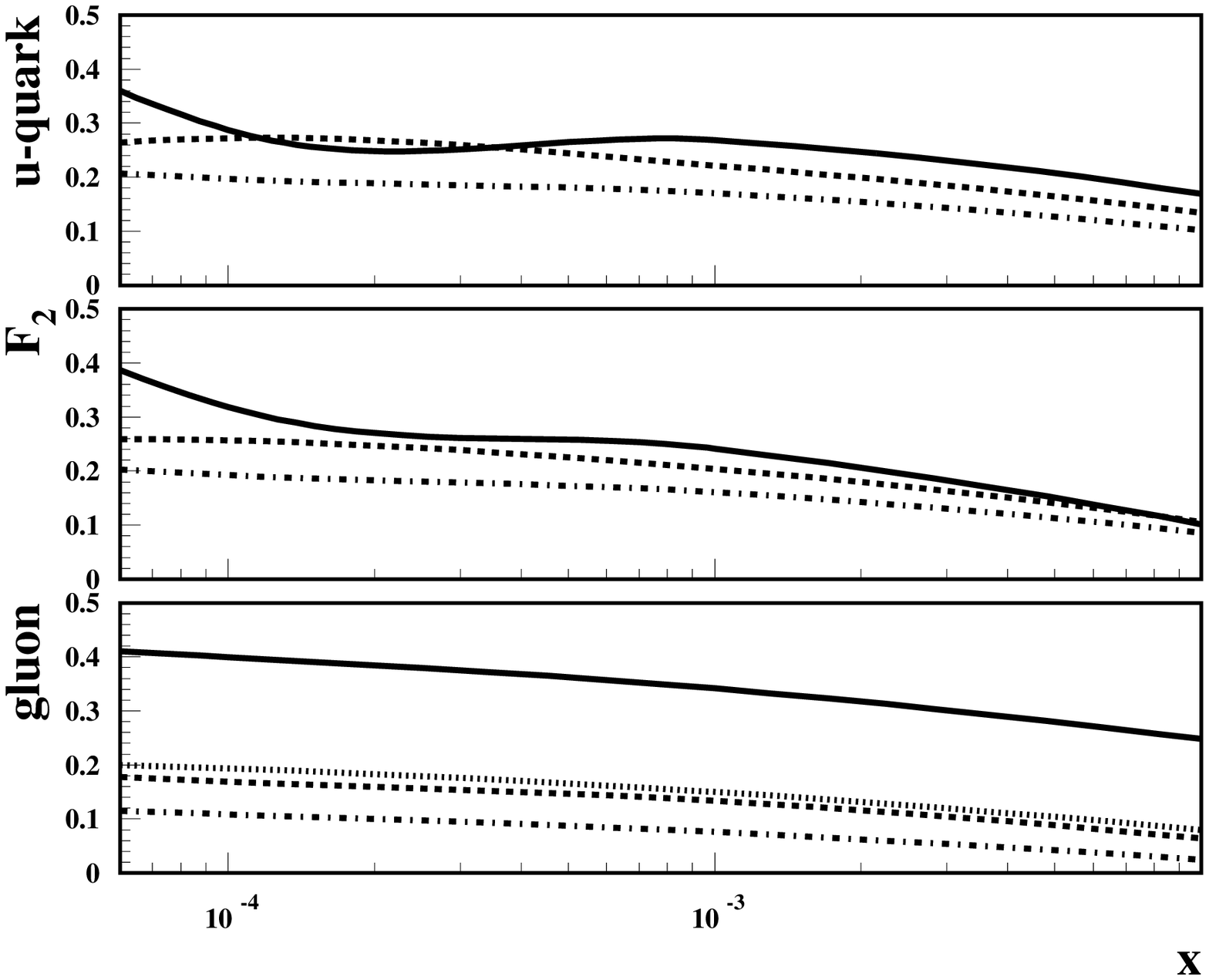,width=13cm,height=13cm}
\vskip 0cm
\caption{The ratios $f_{j/A}^{D(2)}/f_{j/A}$ for the $u$-quarks and gluons and 
NLO $F_{2A}^{D(2)}/F_{2A}$ for $^{208}$Pb. The solid curves correspond to $Q=2$ GeV; the dashed 
curves correspond to $Q=10$ GeV; the dot-dashed curves correspond to $Q=100$ GeV.
In addition, for the gluons the dotted curve correspond to $Q=5$ GeV.
}
\label{fig:pb}
\end{center}
\end{figure}

Having obtained $f_{j/A}^{D(3)}(x,Q^2,x_{\Pomeron})$, they can be integrated over 
over $x_{\Pomeron}$ at fixed Bjorken $x$, just like in Eq.~(\ref{eq:d2}). The resulting   
$f_{j/A}^{D(2)}/f_{j/A}$ ratios for the $u$-quarks and gluons and NLO
$F_{2A}^{D(2)}/F_{2A}$ are presented in Fig.~\ref{fig:ca40} for $^{40}$Ca
and in Fig.~\ref{fig:pb} for $^{208}$Pb. 
From Figs.~\ref{fig:ca40} and ~\ref{fig:pb} one can see that the fraction of the
 diffractive events in DIS at small $x$ for moderately heavy and heavy nuclei is
of the order of $30\%$ and weakly changes with $Q^2$, which is in a good agreement
 with the early estimates of Ref.~\cite{FS96}. 
In the case of gluon-induced reactions, the probability decreases rather significantly
with an increase of $Q^2$. However, the probability still remains at the level of 15-20\%
 at $Q=10$ GeV and, hence, it would be
feasible to study this in  ultraperipheral collisions at the LHC using, for
instance, production
of heavy flavors similarly to the case of inclusive production considered in~\cite{Vogt}.
Another option  is to  use dijet production
like it was done in the proton case in the ZEUS~\cite{dijetZEUS} and 
H1~\cite{H1:1997} experiments.

Another conclusion that can be drawn from 
Figs.~\ref{fig:ca40} and ~\ref{fig:pb} is
that the $A$-dependence of the probability
 of coherent diffraction is rather weak
for $A \geq 40$. For these values of $A$, the interaction for 
the central impact parameters is close to
being completely absorptive (black) with a 
small contribution from the opaque nuclear edge.
Moreover, the $A$-dependence is weaker in the gluon case since the gluon
 interactions at the $Q_0$ scale are already 
close to the black limit, even for the nucleon.

Mathematically this pattern is a result of
a compensation of two effects --  
stronger small-$x$ nuclear shadowing in the case
of coherent diffraction compared to the inclusive case,
 is compensated by the nuclear
form factor, as a consequence of nuclear coherence.

It is worth noting a qualitative difference
 between the $A$-dependence of the fraction
of the diffractive events in the quark and 
gluon-induced processes at small $x$.
 In the gluon case,
it is a very weak function of $A$  because already in the proton case, 
the probability of diffraction is close to one half,
the maximal value allowed by unitarity. At the same time in the 
quark case a steady growth with $A$ 
is predicted since for the proton the probability
 of diffraction in this channel 
is rather small and, hence, the increase of the blackness of 
the interaction with $A$
leads  to a gradual increase of the diffraction probability
to the values close enough to the black body limit.

\section{Conclusions and Discussion}
\label{sec:conclusions}

We study small-$x$ coherent diffraction in DIS on nuclear targets 
using the theory of 
leading twist nuclear shadowing and the QCD factorization theorem
 for hard diffraction.
 It is demonstrated that Bjorken $x$ and $A$-dependent nuclear
shadowing explicitly breaks down Regge factorization in diffraction,
 which means that at
fixed $x_{\Pomeron}$, nuclear parton distribution functions depend not only on
 $\beta=x/x_{\Pomeron}$
but also on Bjorken $x$ and $A$.

We calculate nuclear DPDFs (Figs.~\ref{fig:ca40_beta_new}, \ref{fig:pb_beta_new} and 
\ref{fig:largebeta})
 as functions of $\beta$, $Q^2$, $x_{\Pomeron}$ and the atomic number $A$.
Like in the free nucleon case,
the gluon nDPDF is much larger than the quark nDPDF. Using the calculated nDPDFs, 
the contribution of coherent diffraction to the total probability is estimated for
the $u$-quark and gluon channels as well as for the NLO $F_2$ structure functions 
(Figs.~\ref{fig:ca40} and \ref{fig:pb}). The key result is an observation of dramatically
different patterns of the $A$-dependence. In the quark channel and in the $F_2$ case,
the probability of diffraction increases with $A$,
reaching
about $30\%$ at $x \sim 10^{-4}$ and $Q_0^2 \sim 4$ GeV$^2$ and for $A \sim 200$.
 In the gluon channel, the 
probability of diffraction is already large for the proton and, hence, it changes (decreases)
rather insignificantly when the proton target is replaced by the heavy nuclear target: 
the probability remains at the level of $\sim 40$\% at $x \sim 10^{-4}$ and
  $Q_0^2 \sim 4$ GeV$^2$.

The $Q^2$-dependence of the probability of diffraction is also  different in the 
quark and the gluon channels. The very large gluon diffractive distribution makes QCD evolution
of the ratios $u_{A}^{D(2)}/u_{A}$  and NLO $F_{2A}^{D(2)}/F_{2A}$ rather weak. At the same time,
the ratio $g_{A}^{D(2)}/g_{A}$ falls off rapidly as $Q^2$ increases (compare the solid and 
broken curves in Figs.~\ref{fig:ca40} and \ref{fig:pb}).

From the experimental point of view, coherent diffraction in deep 
inelastic scattering on nuclei can be identified via a two-step
procedure.
First, similarly to the case of $ep$ scattering, one selects events with
a rapidity gap. Second, one needs to separate the coherent and
incoherent diffraction. This can be readily done
using the lack of neutrons in the zero angle
neutron calorimeter since the break-up of the nucleus in
 incoherent diffraction results in production
of  several evaporation neutrons, see discussion in~\cite{Krasny}. 
 In addition,   the ratio of
incoherent diffraction to coherent diffraction is expected to be
$\sim 0.1-0.15$~\cite{Krakow}. Hence overall in the collider kinematics
the   task of selecting the diffractive channel
without
break-up of the target appears to be much easier in the nucleus case
than in the proton case.
The 
$t$-dependence of coherent diffraction
originates primarely from
 the factor $\left(F_A(t)\right)^2$ where $F_A(t)$ is the nuclear form factor.
Hence average $t$ are small and it is hardly possible to measure
the $t$-dependence of the diffractive amplitude for the case of the
large masses of the produced diffractive system. However 
since the $t$-dependence is mostly trivial, inability to measure the
differential cross section would not lead to a 
significant loss of information about the dynamics of diffraction.
Note also that that the break-up channel originates mostly due to the
scattering off the edge of the nucleus, leading to the same pattern of
diffraction as in the scattering off a free nucleon.  Hence we predict
a different $\beta,x_{\Pomeron}, Q^2$ dependence of the hard
diffraction
in incoherent and coherent diffraction.

Nuclear diffractive PDFs, discussed in this paper, exhibit novel 
effects, which are large enough to be 
  measured in the ultraperipheral collisions at the LHC.
We also would like to emphasize that  the proximity  of
the probability of hard diffraction to the unitarity limit at
$Q^2 \sim 4$ GeV$^2$ shows that the color transparency phenomenon and 
related to Bjorken scaling decomposition
over powers of $1/Q^{2}$, which are typical for DIS,
disappear in the vicinity of these $Q^2$.
Thus, nuclear shadowing  does not preclude observation
of a  variety of phenomena characteristic for the unitarity limit
for the gluon channel in the case of the nucleon and  nuclear targets at $x \le
10^{-3}$, and for  the quark channel for $ x \leq 10^{-4}$ for heavy
nuclear targets. In particular, we expect that blackening
of the interaction will reveal itself in  heavy ion collisions at the LHC
(and to less extent at RHIC) in the filtering out of nonperturbative QCD
effects and producing a pQCD phase in the proton-nucleus collisions
in the proton fragmentation region~\cite{Dumitru:2002wd} and in the heavy ion
 collisions  in the ion
fragmentation regions~\cite{Frankfurt:2002js}.

Numerical results presented in this paper are available 
from V. Guzey (vadim.guzey@tp2.ruhr-uni-bochum.de) 
upon request.

This work was supported by Sofia Kovalevskaya 
Program of the Alexander von Humboldt  Foundation (Germany) and Department of Energy (USA)
and GIF.

\end{document}